\documentclass[onecolumn,showpacs,preprintnumbers]{revtex4}
\usepackage{mathrsfs}
\usepackage{graphicx}
\usepackage{dcolumn}
\usepackage{bm}
\usepackage{epsfig}
\usepackage{amsmath,amssymb}

\begin{document}
\title{Analysis of the strong vertices of $\Sigma_cND^{*}$ and $\Sigma_bNB^{*}$ in QCD sum rules}
\author{Guo-Liang Yu$^{1}$}
\email{yuguoliang2011@163.com}
\author{Rong-Hua Guan$^{1}$}
\author{Zhi-Gang Wang$^{1}$}
\email{zgwang@aliyun.com}

\affiliation{$^1$ Department of Mathematics and Physics, North China
Electric power university, Baoding 071003, People's Republic of
China}
\date{\today }

\begin{abstract}
The strong coupling constant is an important parameter which can
help us to understand the strong decay behaviors of baryons. In our
previous work, we have analyzed strong vertices $\Sigma_{c}^{*}ND$,
$\Sigma_{b}^{*}NB$, $\Sigma_{c}ND$, $\Sigma_{b}NB$ in QCD sum rules.
Following these work, we further analyze the strong vertices
$\Sigma_{c}ND^{*}$ and $\Sigma_{b}NB^{*}$ using the three-point QCD
sum rules under Dirac structures $q\!\!\!/p\!\!\!/\gamma_{\alpha}$
and $q\!\!\!/p\!\!\!/p_{\alpha}$. In this work, we first calculate
strong form factors considering contributions of the perturbative
part and the condensate terms $\langle\overline{q}q\rangle$,
$\langle\frac{\alpha_{s}}{\pi}GG\rangle$ and
$\langle\overline{q}g_{s}\sigma Gq\rangle$. Then, these form factors
are used to fit into analytical functions. According to these
functions, we finally determine the values of the strong coupling
constants for these two vertices $\Sigma_{c}ND^{*}$ and
$\Sigma_{b}NB^{*}$.
\end{abstract}


\maketitle

\begin{large}
\textbf{1 Introduction}
\end{large}

In the past 20 years, we have witnessed the baryon spectrum been
established step by step with the cooperative efforts from both
experimentalists and theorists. Up to now, about 20 charmed baryon
candidates have been discovered by different experimental
collaborations\cite{Patrignani}. Besides, many bottom baryons, e.g.
$\Lambda_{b}$, $\Xi_{b}$, $\Sigma_{b}$, $\Sigma^{*}_{b}$, and
$\Omega_{b}$, have also been announced by CDF and LHCb
collaborations\cite{Basile,Aaltonen,Aaij1,Aaij2}. In 2017, LHCb
Collaboration reported the observation of the doubly charmed baryon
$\Xi_{cc}^{++}$ in the $\Lambda_{c}^{+}K^{-}\pi^{+} \pi^{+}$ mass
spectrum\cite{Aaij3}, which has became a new motivation for
researchers to devote themselves to studying the properties of these
heavy baryons.

These charmed and bottom baryons, which contain at least a heavy
quark, provide a unique system for testing models of quantum
chromodynamics (QCD), the theory that describes the strong
interaction. In other words, these special baryons can be looked as
a particular laboratory for studying dynamics between light quarks
and heavy ones, and also as an excellent ground for testing validity
of the quark model and heavy quark symmetry. The properties of these
baryons such as the mass spectrum, the magnetic moments, the strong,
electromagnetic and weak decay behaviors have been studied with a
variety of theoretical
models\cite{Faes,Pate,LiuX,ChenHX1,Chun,ChenW,Zhjr,Karl1,Karl2,Nava,Khod,Alie,Azi1,Azi2,Wzg,KangXW,Esposito}.
As an important parameter, the strong coupling constant can not only
help us to know about the strong decay behaviors of baryons but also
play an essential role for understanding its inner structure. Thus,
people calculated some of the strong coupling constants
$g_{\Omega^{*}_{c}\Omega^{*}_{c}\phi}$,
$g_{\Omega^{*}_{c}\Xi^{*}_{c}K^{*}}$,
$g_{\Xi^{*}_{c}\Sigma^{*}_{c}K^{*}}$,
$g_{\Omega^{*}_{b}\Omega^{*}_{b}\phi}$,
$g_{\Xi^{*}_{b}\Sigma^{*}_{b}K^{*}}$, $g_{\Xi^{*}_{b}\Xi_{b}\pi}$,
$g_{\Xi^{*}_{c}\Xi_{c}\pi}$, $g_{\Lambda_{b}NB^{*}}$,
$g_{\Lambda_{b}N^{*}B^{*}}$, $g_{\Lambda_{c}ND^{*}}$,
$g_{\Lambda_{c}N^{*}D^{*}}$, $g_{\Sigma_{b}N^{*}B^{*}}$ and
$g_{\Sigma_{c}N^{*}D^{*}}$,
\emph{etc}\cite{Azi1,Azi2,Wzg5,Navar,Khodj,GuoLY2,GuoLY4,Azizi4}.

To calculate the strong coupling constant, we can adopt several
theoretical models including perturbative and non-perturbative
methods. The QCD sum rules, proposed by Shifman, Vainshtein, and
Zakharov\cite{Shifman}, connects hadron properties and QCD
parameters\cite{Reind}. It has been widely used to study the
properties of the
hadrons\cite{Ioffe,Belyaev,Brac1,Brac2,Alie2,Alie3,Alie4,Doi,Altm,Wzg3,Cerq,Rodr,Yazi,Khos1,Khos2,Rein,Pasc}.
In our previous work, we have analyzed the strong vertices
$\Sigma_{c}^{*}ND$, $\Sigma_{b}^{*}NB$, $\Sigma_{c}ND$ and
$\Sigma_{b}NB$ in QCD sum rule framework\cite{GuoLY2,GuoLY4}. As a
continuation of these work, we analyze the strong vertices
$\Sigma_{c}ND^{*}$ and $\Sigma_{b}NB^{*}$ using the three-point QCD
sum rules under the Dirac structures
$q\!\!\!/p\!\!\!/\gamma_{\alpha}$ and $q\!\!\!/p\!\!\!/p_{\alpha}$.
This paper is organized as follows. After the Introduction, we
present details of the analysis of vertices $\Sigma_{c}ND^{*}$ and
$\Sigma_{b}NB^{*}$. In Sec.3, we present the numerical results and
discussions. Finally, the paper ends with the Conclusion.

\begin{large}
\textbf{2 QCD sum rules for $\Sigma_{c}ND^{*}$ and
 $\Sigma_{b}NB^{*}$}
\end{large}

In order to obtain the strong coupling constants of vertices
$\Sigma_{c}ND^{*}$ and $\Sigma_{b}NB^{*}$, we write out the
following three-point correlation function,
\begin{eqnarray}
\Pi_{\alpha}(p,p',q)=i^{2}\int d^{4}x\int
d^{4}ye^{-ip.x}e^{ip'.y}\Big\langle0|\mathcal
{T}\Big(J_{N}(y)J_{D^{*}[B^{*}]}^{\alpha}(0)\overline{J}_{\Sigma_{c}[\Sigma_{b}]}(x)\Big)
|0\Big\rangle,
\end{eqnarray}
where $J_{N}$, $J^{\alpha}_{D^{*}[B^{*}]}$ and
$J_{\Sigma_{c}[\Sigma_{b}]}$ denote interpolating currents of $N$,
$D^{*}[B^{*}]$ and $\Sigma_{c}[\Sigma_{b}]$, and $\mathcal {T}$ is
the time ordered product. Currents are composite operators made of
quark and gluon fields that can create the studied hadrons from
vacuum. It has the same quantum numbers with these
hadrons\cite{Ioffe,Cola}. In this paper, the interpolating currents
are written as,
\begin{eqnarray}
\notag
&& J_{\Sigma_{c}[\sum_{b}]}(x)=\epsilon_{ijk}\Big(u^{iT}(x)C\gamma_{\mu}d^{j}(x)\Big)\gamma_{5}\gamma^{\mu}c[b]^{k}(x) \\
\notag
&& J_{N}(y)=\epsilon_{ijk}\Big(u^{iT}(y)C\gamma_{\mu}u^{j}(y)\Big)\gamma_{5}\gamma^{\mu}d^{k}(y)\\
&&
J^{\alpha}_{D^{*}[B^{*}]}(0)=\overline{u}(0)\gamma_{\alpha}c[b](0)
\end{eqnarray}

In QCD sum rule framework, there is a region of $p$ where
correlation function can be equivalently described at both hadron
and quark sector. The former is called the phenomenological side and
the latter is called QCD or operator product expansion(OPE) side.
Matching these two sides of the sum rule, we can obtain information
about hadron properties.

\begin{large}
\textbf{2.1 The phenomenological side}
\end{large}

On the phenomenological side, we insert a complete set of
intermediate hadronic states into the correlation
$\Pi_{\alpha}(p,p',q)$. These intermediate states have the same
quantum numbers as the current operators $J_{N}$, $J_{D^{*}[B^{*}]}$
and $J_{\Sigma_{c}[\Sigma_{b}]}$. Isolation of ground-state
contributions results in the following expression,
\begin{eqnarray}
 \notag
 \Pi^{HAD}_{\alpha}(p,p',q)=&&\frac{\Big \langle 0| J_{N}|N(p')\Big
\rangle \Big \langle 0|
J^{\alpha}_{D^{*}[B^{*}]}|D^{*}[B^{*}](q)\Big \rangle \Big \langle
\Sigma_{c}[\Sigma_{b}](p)|
\overline{J}_{\Sigma_{c}[\Sigma_{b}]}|0\Big
\rangle}{(p^{2}-m^{2}_{\Sigma_{c}[\Sigma_{b}]})(p'^{2}-m^{2}_{N})(q^{2}-m^{2}_{D^{*}[B^{*}]})}\\
&& \Big\langle
N(p')D^{*}[B^{*}](q)|\Sigma_{c}[\Sigma_{b}](p)\Big\rangle+\cdots
\end{eqnarray}
Here, ellipsis denotes the contributions from higher resonances and
continuum states. We substitute the matrix elements appearing in
Eq.(3) with the following parameterized equations,
\begin{eqnarray}
\notag
&&\langle 0| J_{N}|N(p') \rangle=\lambda_{N}u_{N}(p',s'),\\
\notag
&&\langle0|J^{\alpha}_{D^{*}[B^{*}]}|D^{*}[B^{*}](q) \rangle=m_{D^{*}[B^{*}]}f_{D^{*}[B^{*}]}\varepsilon^{*}_{\alpha},\\
\notag &&\langle
\Sigma_{c}[\Sigma_{b}](p)|\overline{J}_{\mu\Sigma_{c}[\Sigma_{b}]}|0
\rangle=\lambda_{\Sigma_{c}[\Sigma_{b}]}\overline{u}_{\mu\Sigma_{c}[\Sigma_{b}]}(p,s),\\
&&\langle N(p')D^{*}[B^{*}](q)|\Sigma_{c}[\Sigma_{b}](p)\rangle=
\varepsilon_{\beta}\overline{u}_{N}(p',s')\big[g_{1}\gamma_{\beta}-g_{2}\frac{i\sigma_{\beta\nu}}{m_{\Sigma_{c}[\Sigma_{b}]}+m_{N}}q^{\nu}\big]u_{\Sigma_{c}[\Sigma_{b}]}(p,s).
\end{eqnarray}
In the hadron degrees of freedom, the correlation function
$\Pi_{\alpha}(p,p',q)$ is finally decomposed into the following
different dirac structures,
\begin{eqnarray}
\notag
\Pi^{HAD}_{\alpha}(p,p',q)=&&\frac{m_{D^{*}[B^{*}]}f_{D^{*}[B^{*}]}\lambda_{N}\lambda_{\Sigma_{c}[\Sigma_{b}]}}{(p^{2}-m^{2}_{\Sigma_{c}[\Sigma_{b}]})(p'^{2}-m^{2}_{N})(q^{2}-m^{2}_{D^{*}[B^{*}]})}\\
 \notag
&&\times
\Big\{(g_{1}+g_{2})(m_{N}-m_{\Sigma_{c}[\Sigma_{b}]})p\!\!\!/\gamma_{\alpha}+\frac{(g_{1}+g_{2})m_{\Sigma_{c}[\Sigma_{b}]}(m_{N}+m_{\Sigma_{c}[\Sigma_{b}]})}{}q\!\!\!/\gamma_{\alpha}
\\
 \notag && -(g_{1}+g_{2})q\!\!\!/p\!\!\!/\gamma_{\alpha}+ (g_{1}+g_{2})m_{\Sigma_{c}[\Sigma_{b}]}(m_{\Sigma_{c}[\Sigma_{b}]}-m_{N})\gamma_{\alpha}\\
 \notag &&-2g_{1}p\!\!\!/p_{\alpha}+2\big[g_{1}+g_{2}\frac{m_{N}}{m_{N}+m_{\Sigma_{c}[\Sigma_{b}]}}\big]q\!\!\!/p_{\alpha}-\frac{2g_{2}}{m_{N}+m_{\Sigma_{c}[\Sigma_{b}]}}q\!\!\!/p\!\!\!/p_{\alpha}\\
\notag && -2\big[g_{1}m_{N}+g_{2}(m_{N}-m_{\Sigma_{c}[\Sigma_{b}]})\big]p_{\alpha}\\
\notag &&+\big[g_{1}\frac{m^{2}_{\Sigma_{c}[\Sigma_{b}]}-m^{2}_{N}}{q^{2}}-g_{2}\big]p\!\!\!/q_{\alpha}+\big[g_{1}\frac{m_{\Sigma_{c}[\Sigma_{b}]}(m_{N}-m_{\Sigma_{c}[\Sigma_{b}]})}{q^{2}}+g_{2}\frac{m_{N}}{m_{N}+m_{\Sigma_{c}[\Sigma_{b}]}}\big]q\!\!\!/q_{\alpha}\\
\notag && +\big[g_{1}\frac{m_{N}-m_{\Sigma_{c}[\Sigma_{b}]}}{q^{2}}+\frac{g_{2}}{m_{N}+m_{\Sigma_{c}[\Sigma_{b}]}}\big]q\!\!\!/p\!\!\!/q_{\alpha}  \\
\notag && +\big[(2g_{1}+g_{2})m_{\Sigma_{c}[\Sigma_{b}]}+\frac{g_{1}m_{\Sigma_{c}[\Sigma_{b}]}(m^{2}_{\Sigma_{c}[\Sigma_{b}]}-m^{2}_{N})}{q^{2}}\big]q_{\alpha}+...\\
\end{eqnarray}

If all criteria of the QCD sum rules are satisfied, each dirac
structure in Eq.(5) can be used to carry out the calculation. It is
true that people indeed had different choice about this problem in
the similar work\cite{Azzi6,Azi1,WangZG5}. And these researches
indicated that different Dirac structures can really lead to
compatible results. For simplicity, we choose
$q\!\!\!/p\!\!\!/\gamma_{\alpha}$ and $q\!\!\!/p\!\!\!/p_{\alpha}$
Dirac structures to perform our analysis.

\begin{large}
\textbf{2.2 The OPE side}
\end{large}

Considering all possible contractions of the quark fields with
Wick's theorem, we write the correlation function as follows,
\begin{eqnarray}
\notag\ \Pi_{\alpha}^{OPE}(p,p',q)=&&i^{2}\int d^{4}x\int
d^{4}ye^{-ip.x}e^{ip'.y}\epsilon_{abc}\epsilon_{ijk}\\
&& \notag
\times\Big\{\gamma_{5}\gamma_{\mu}S_{d}^{cj}(y-x)\gamma_{\nu}CS_{u}^{biT}(y-x)C\gamma_{\mu}S_{u}^{ah}(y)\gamma_{\alpha}S_{c[b]}^{hk}(-x)\gamma_{\nu}\gamma_{5}\\
&&-\gamma_{5}\gamma_{\mu}S_{d}^{cj}(y-x)\gamma_{\nu}CS_{u}^{aiT}(y-x)C\gamma_{\mu}S_{u}^{bh}(y)\gamma_{\alpha}S_{c[b]}^{hk}(-x)\gamma_{\nu}\gamma_{5}\Big\}
\end{eqnarray}
Here, $S_{q[Q]}$ stands for up- and down-quark, or charm- and
bottom-quark propagators which will be replaced by the following
propagators\cite{Pasc,Rein}.
\begin{eqnarray}
 S_{u[d]}^{mn}(x)=&&i\frac{x\!\!\!/}{2\pi^{2}x^{4}}\delta_{mn}-\frac{m_{u[d]}}{4\pi^{2}x^{2}}\delta_{mn}-\frac{\langle
 \overline{q}q\rangle}{12}\Big(1-i\frac{m_{u[d]}}{4}x\!\!\!/\Big)-\frac{x^{2}}{192}m_{0}^{2}\langle
 \overline{q}q\rangle\Big( 1-i\frac{m_{u[d]}}{6}x\!\!\!/\Big)\\
 &&
 \notag-\frac{ig_{s}\lambda_{A}^{ij}G^{A}_{\theta\eta}}{32\pi^{2}x^{2}}\Big[x\!\!\!/\sigma^{\theta\eta}+\sigma^{\theta\eta}x\!\!\!/\Big]+\cdots,
\end{eqnarray}
\begin{eqnarray}
\notag
 S_{c[b]}^{mn}(x)=&&\frac{i}{(2\pi)^{4}}\int
 d^{4}ke^{-ik.x}\Big\{\frac{\delta_{mn}}{k\!\!\!/-m_{c[b]}}-\frac{g_{s}G_{mn}^{\alpha\beta}}{4}\frac{\sigma_{\alpha\beta}(k\!\!\!/+m_{c[b]})+(k\!\!\!/+m_{c[b]})\sigma_{\alpha\beta}}{(k^{2}-m_{c[b]}^{2})^{2}}\\
 &&+\frac{\pi^{2}}{3}\Big\langle\frac{\alpha_{s}GG}{\pi}\Big\rangle\delta_{mn}m_{c[b]}\frac{k^{2}+m_{c[b]}k\!\!\!/}{(k^{2}-m_{c[b]}^{2})^{4}}+\cdots\Big\}
\end{eqnarray}
After a lengthy derivation, which need us to carry out the process
of Fourier transformation, Feynman parametrization $etc$, we can
obtain the same Dirac structures as the phenomenological side in
Eq.$(5)$. One can consult reference\cite{GuoLY2} for technical
details of these processes. For each Dirac structure, the
correlation function can be decomposed into two parts, perturbative
term and non-perturbative term,
\begin{eqnarray}
\Pi^{OPE}_{i}=\Pi^{pert}_{i}+\Pi^{non-pert}_{i}
\end{eqnarray}
where the latter is composed of condensate terms, and $i$ stands for
different Dirac structure. Using dispersion relation, the
correlation function for a special Dirac structure can be written
as,
\begin{eqnarray}
\Pi^{OPE}_{i}(q^{2})=\int_{s_{1}}^{s_{0}} ds\int_{u_{1}}^{u_{0}}
du\frac{\rho_{i}^{pert}(s,u,q^{2})+\rho_{i}^{non-pert}(s,u,q^{2})}{(s-p^{2})(u-p'^{2})}
\end{eqnarray}
Here, $\rho_{i}^{pert[non-pert]}$ is spectral density which is
obtained from the imaginary part of correlation function. During
these derivations, we set $s=p^2$, $u=p'^2$ and $q=p-p'$ in the
spectral densities. For dirac structures
$q\!\!\!/p\!\!\!/\gamma_{\alpha}$ and $q\!\!\!/p\!\!\!/p_{\alpha}$,
its perturbative term are written as,
\begin{eqnarray}
\notag\
\rho_{q\!\!\!/p\!\!\!/\gamma_{\alpha}}^{pert}(s,u,q^2)&&=\frac{3}{32\pi^{4}}\int^{1}_{0}dx\int^{1-x}_{0}\Big\{\frac{m_{u}\big[m_{b}x-m_{u}y\big]-m_{d}\big[m_{b}x+m_{u}(2x+y-2)\big]}{x+y-1} \\
&& \times\Theta[H(s,u,q^{2})]\Big\}\ dy,
\end{eqnarray}
\begin{eqnarray}
\notag\
\rho_{q\!\!\!/p\!\!\!/p_{\alpha}}^{pert}(s,u,q^2)&&=\frac{3}{32\pi^{4}}\int^{1}_{0}dx\int^{1-x}_{0}\Big\{\frac{2\big[m_{b}x(x+y)+y\big(m_{u}(2x+2y-3)-3m_{d}(x+y-1)\big)\big]}{x+y-1}dy\\
&& \times\Theta[H(s,u,q^{2})]\Big\}\ dy,
\end{eqnarray}
where $H[s,u,q^2]=x (m_{c[b]}^2-q^2 y)+s x (x+y-1)+uy(x+y-1)$ and
$\Theta$ stands for a unit-step function. Considering these limits,
the integral limit for parameter $y$ can be explicitly expressed as
$0\leq y\leq min\Big\{1-x,y_{1}\Big\}$, where
$y_{1}=\frac{[(s+u-q^2)-u]-\sqrt{\Delta}}{-2u}$ and
$\Delta=\Big[(s+u-q^2)^2-4su\Big]x^2-2u\Big[(s+u-q^2)+2m_{c[b]}^{2}\Big]x+4su+u^{2}$

For non-perturbative terms, its spectral densities are written as,
\begin{eqnarray}
\notag\ &&
\rho_{q\!\!\!/p\!\!\!/\gamma_{\alpha}}^{non-pert}(s,u,q^2)\\ \notag\
&& =\frac{\langle\overline{q}q\rangle}{64\pi^2}\int^{1}_{0}\Big\{\frac{-16\big[(m_{d}+m_{u})(x+y_{1}-1)\big]}{u\big(x+2y_{1}-1\big)+\big(s-q^{2}\big)x} \Big\}dx \\
\notag\
&&-\frac{\langle\overline{q}q\rangle}{16\pi^2}\Big\{\frac{\big[m_{d}-m_{u}\big]\big[m_{b}(m_{u}-2m_{b})+2q^{2}\big]}{(m_{b}^{2}-q^{2})^2}\Theta[u]-\frac{2m_{d}m_{u}^{2}}{m_{b}^{2}-q^{2}}\delta[u]\Big\}\\
\notag\
&&-\frac{\langle \alpha_{s}\frac{G^{2}}{\pi}\rangle}{16\times64\pi^2}\int^{1}_{0}\Big\{\frac{4um_{b}\big(m_{d}-m_{u}\big)\big(3x^3-2x^2\big)x}{\sqrt{\Delta}\big[u\big(x+2y_{1}-1\big)+\big(s-q^{2}\big)x\big]^2}\Big\}dx\\
\notag\ &&
-\frac{m_{0}^{2}\langle\overline{q}q\rangle}{64\times4\pi^{2}}\Big\{\frac{8\Big\{m_{b}^{3}\big[m_{b}(6m_{d}-3m_{u})+2m_{u}(m_{u}-m_{d})\big]+4m_{b}^{2}q^{2}\big[m_{u}-2m_{d}\big]+q^{4}[2m_{d}-m_{u}\big]\Big\}}{\big(m_{b}^{2}-q^{2}\big)^{4}}\Big\}\Theta[u]\\
\notag\ &&
-\frac{m_{0}^{2}\langle\overline{q}q\rangle}{64\times4\pi^{2}}\Big\{\frac{8\big[(m_{b}^{2}-q^{2})\big(6m_{d}-3m_{u})+2m_{d}m_{u}^{2}+m_{b}m_{u}(m_{u}-m_{d})]}{3\big(m_{b}^{2}-q^{2}\big)^{3}}\Big\}\Theta[u]\\
\notag\
&&-\frac{m_{0}^{2}\langle\overline{q}q\rangle}{64\times4\pi^{2}}\Big\{\frac{8\Big\{9m_{b}^{4}\big[2m_{d}-m_{u}\big]+3m_{b}^{3}m_{u}\big[m_{u}-m_{d}\big]+m_{b}m_{u}\big[m_{d}-m_{u}\big]\big[2s+q^{2}\big]\Big\}}{3\big(m_{b}^{2}-q^{2}\big)^{3}}\\
&&-\frac{8\Big\{q^{2}\big[2m_{d}(m_{u}^{2}-3s-6q^{2})+3m_{u}(s+2q^{2})\big]-3m_{b}^{2}\big[2m_{d}(m_{u}^{2}-s-5q^{2})+m_{u}(s+5q^{2})\big]\Big\}}{3\big(m_{b}^{2}-q^{2}\big)^{3}}\Big\}\delta[u]
\end{eqnarray}
\begin{eqnarray}
\notag\ && \rho_{q\!\!\!/p\!\!\!/p_{\alpha}}^{non-pert}(s,u,q^2) \\
\notag\
&&=-\frac{\langle\overline{q}q\rangle}{16\pi^2}\frac{2m_{u}\big[m_{u}-m_{d}\big]}{m_{b}^{2}-q^{2}}\delta[u]
-\frac{\langle\alpha_{s}\frac{G^{2}}{\pi}\rangle}{16\times64\pi^2}\int^{1}_{0}\frac{m_{b}}{3\big[x+y_{1}-1\big]\big[u\big(x+2y_{1}-1\big)+\big(s-q^{2}\big)x\big]\sqrt{\Delta}}\\
\notag\ && \times\Big\{
\frac{4ux^{2}\big[-5x^{2}+x(4-5y_{1})+2y_{1}\big]}{u(x+2y_{1}-1)+(s-q^{2})x}+\frac{2x^{2}\big[-5x^{2}+x(4-5y_{1})+2y_{1}\big]}{x+y_{1}-1}+2x^{2}\big[5x-2\big]\Big\}dx\\
\notag\
&&+\frac{m_{0}^{2}\langle\overline{q}q\rangle}{64\times4\pi^{2}}\Big\{\frac{16m_{b}^{3}\big(3m_{b}-2m_{u}\big)-4m_{b}^{2}q^{2}+q^{4}}{3\big(m_{b}^{2}-q^{2}\big)^{4}}+\frac{16\big(3m_{b}^{2}-2m_{b}m_{u}-3q^{2}\big)}{9\big(m_{b}^{2}-q^{2}\big)^{3}}\Big\}\Theta[u]\\
\notag\
&&+\frac{m_{0}^{2}\langle\overline{q}q\rangle}{64\times4\pi^{2}}\Big\{\frac{16\big[12m_{b}^{4}-4m_{b}^{3}m_{u}-3m_{b}^{2}\big(4m_{d}m_{u}-4m_{u}^{2}+s+7q^{2}\big)\big]}{9\big(m_{b}^{2}-q^{2}\big)^{3}}\\
&&+\frac{2m_{b}m_{u}\big(s+q^{2}\big)+3q^{2}\big(2m_{d}m_{u}-2m_{u}^{2}+s+3q^{2}\big)}{9\big(m_{b}^{2}-q^{2}\big)^{3}}\Big\}\delta[u]
\end{eqnarray}
where $\delta$ stands for Delta function.

\begin{large}
\textbf{3 The results and discussions}
\end{large}

To calculate strong form factor, we match Eq.(10) with the hadronic
representation Eq.(5), invoking the quark-hadron duality. After
that, we make the change of variables $p^2\rightarrow-P^2$,
$p'^2\rightarrow-P'^2$, $q^2\rightarrow-Q^2$ and perform a double
Borel transformation in $P^2$, $P'^2$, introducing two Borel
parameters $M_1$ and $M_2$. After these preformation, the strong
form factors can be written as,
\begin{eqnarray}
\notag && g_{2\Sigma_{c}ND^{*}[\Sigma_{b}NB^{*}]}(Q^{2})
=\frac{(m_{N}+m_{\Sigma_{c}[\Sigma_{b}]})(Q^2+m^{2}_{D^{*}[B^{*}]})}{2m_{D^{*}[B^{*}]}f_{D^{*}[B^{*}]}\lambda_{N}\lambda_{\Sigma_{c}[\Sigma_{b}]}}e^{\frac{m_{\Sigma_{c}[\Sigma_{b}]}^2}{M_1^2}}e^{\frac{m_N^2}{M_2^2}}
\\
&&\times\int_{(m_{c[b]}+m_{u}+m_{d})^2}^{s_{0}}ds\int_{(2m_{u}+m_{d})^2}^{u_{0}}du
\big[\rho_{q\!\!\!/p\!\!\!/p_{\alpha}}^{pert}(s,u,Q^{2})+\rho_{q\!\!\!/p\!\!\!/p_{\alpha}}^{non-pert}(s,u,Q^{2})\big]e^{-\frac{s}{M_{1}^{2}}}e^{-\frac{u}{M_{2}^{2}}}
\end{eqnarray}
\begin{eqnarray}
\notag &&
g_{1\Sigma_{c}ND^{*}[\Sigma_{b}NB^{*}]}(Q^{2})+g_{2\Sigma_{c}ND^{*}[\Sigma_{b}NB^{*}]}(Q^{2})=\frac{Q^2+m^{2}_{D^{*}[B^{*}]}}{m_{D^{*}[B^{*}]}f_{D^{*}[B^{*}]}\lambda_{N}\lambda_{\Sigma_{c}[\Sigma_{b}]}}e^{\frac{m_{\Sigma_{c}[\Sigma_{b}]}^2}{M_1^2}}e^{\frac{m_N^2}{M_2^2}}\\
&& \times\int_{(m_{c[b]}+m_{u}+m_{d})^2}^{s_{0}}ds
\int_{(2m_{u}+m_{d})^2}^{u_{0}}du
\big[\rho_{q\!\!\!/p\!\!\!/\gamma_{\alpha}}^{pert}(s,u,Q^{2})+\rho_{q\!\!\!/p\!\!\!/\gamma_{\alpha}}^{non-pert}(s,u,Q^{2})\big]e^{-\frac{s}{M_{1}^{2}}}e^{-\frac{u}{M_{2}^{2}}}
\end{eqnarray}

In order to eliminate the contributions from excited and continuum
states at OPE side, two continuum threshold parameters, $s_{0}$ and
$u_{0}$, are adopted as the upper limits of integrals in Eqs.(15)
and (16). Commonly, the values of these parameters are employed as
$s_{0}=(m_{i}+\Delta_{i})^2$ and $u_{0}=(m_{o}+\Delta_{o})^2$, where
$m_{i}$ and $m_{o}$ are ground state masses of the in-coming and
out-coming baryons. In addition, the values of $s_{0}$ and $u_{0}$
in general are expected to be close to the mass squared of the first
excited state of these in-coming and out-coming baryons, which will
lead $\Delta_{i}$ and $\Delta_{o}$ to be about about
$0.3GeV^2\sim0.5GeV^2$. As for the other parameters in Eqs.(15) and
(16), e.g. the masses of the hadrons and the quarks, the decay
constants, the vacuum condensates, their values are all listed in
Table 1.
\begin{table*}[t]
\begin{ruledtabular}\caption{Input parameters used in this analysis.}
\begin{tabular}{c c c c }
  Parameters & \ Values   &\ Parameters &\ Values\\
\hline $m_{\Sigma_{b}}$    &  \   $5811.3\pm1.9$  $MeV$ \cite{Patrignani} &\ $\langle \overline{u}u\rangle$         &  \  $-(0.24\pm0.01)GeV^3$\cite{Ioff2}           \\
$m_{\Sigma_{c}}$     &  \    $2453.98\pm0.16$  $MeV$ \cite{Patrignani} &\ $\langle \overline{d}d\rangle$        &  \  $-(0.24\pm0.01)GeV^3$  \cite{Ioff2}      \\
$m_{N}$       &  \   $ 939.565379 \pm 0.000021$  $MeV$  \cite{Patrignani} &\ $\langle \frac{\alpha_{s}G^{2}}{\pi}\rangle$        &  \    $(0.012\pm0.04)$  $GeV^4$  \cite{Bely}        \\
$m_{D^{*}}$       &  \      $2006.99 \pm 0.15$  $MeV$  \cite{Patrignani}&\  $\lambda_{\Sigma_{b}}$        &  \     $0.062\pm0.018 GeV^3$ \cite{WangZG}   \\
$m_{B^{*}}$      &  \   $5325.2 \pm 0.4$    $MeV$  \cite{Patrignani}&\  $\lambda_{\Sigma_{c}}$        &  \    $0.045\pm0.015 GeV^3 $ \cite{WangZG}   \\
$m_{b}$       &  \     $4.18\pm0.03$ $GeV$ \cite{Patrignani}&\  $\lambda_{N}^{2}$        &  \   $0.0011\pm0.0005$ $GeV^6$   \cite{Azi3}     \\
$m_{c}$          &  \    $1.275 \pm 0.025$   $GeV$ \cite{Patrignani}&\  $f_{B^{*}}$        &  \    $210.3^{+0.1}_{-1.8}$  $MeV$  \cite{Khod4}   \\
$m_{d}$        &  \    $4.8^{+0.5}_{-0.3}$ $MeV$  \cite{Patrignani} &\  $f_{D^{*}}$        &  \    ($241.9^{+10.1}_{-12.1}$)    $MeV$  \cite{Eise}     \\
$m_{u}$        &  \    $2.3^{+0.7}_{-0.5}$ $MeV$   \cite{Patrignani} &\  $m_{0}^{2}$   & \ $0.8\pm0.2GeV^2$\cite{Khod4}\\
\end{tabular}
\end{ruledtabular}
\end{table*}

\begin{figure}[h]
\begin{minipage}[t]{0.45\linewidth}
\centering
\includegraphics[height=5cm,width=7cm]{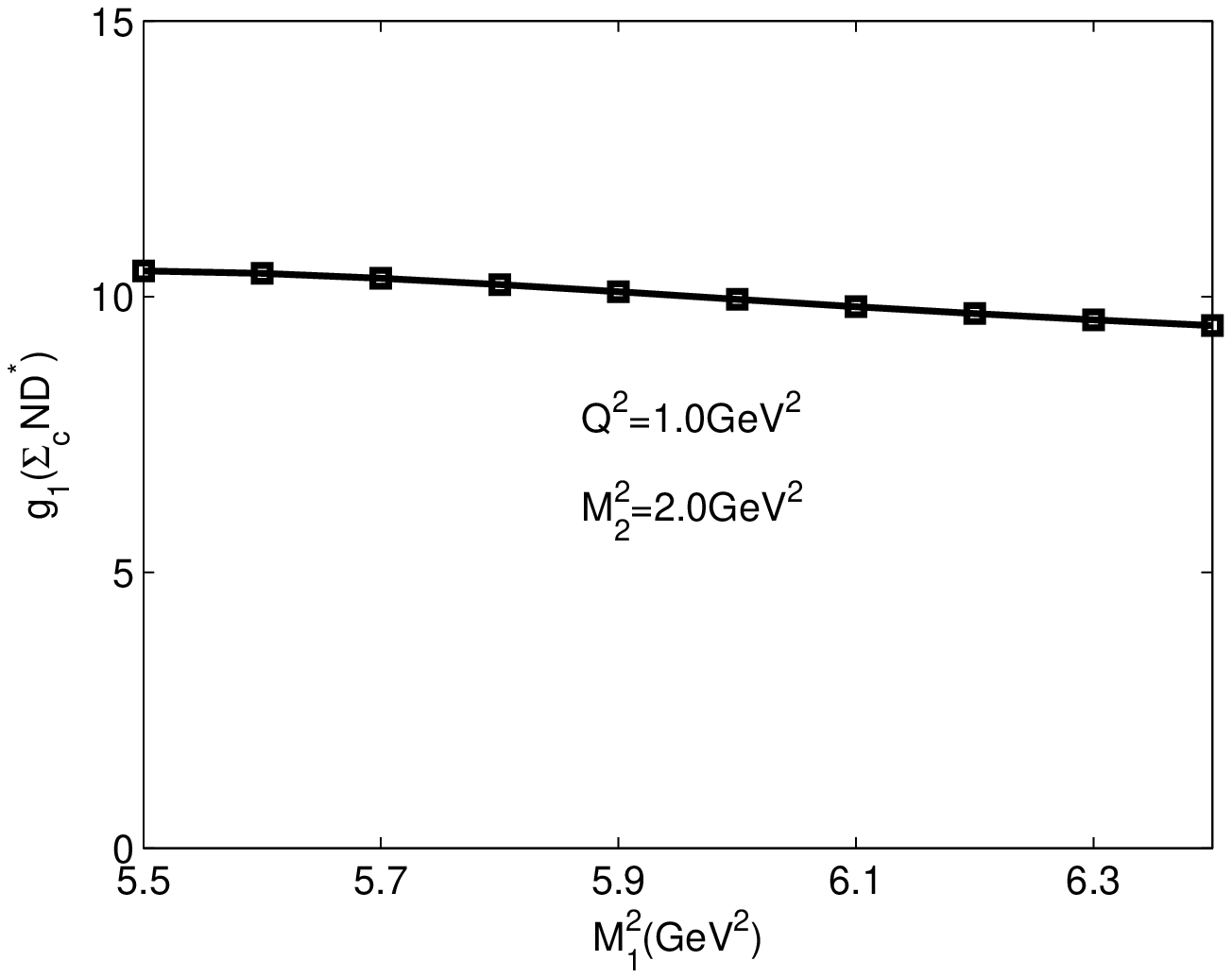}
\caption{The strong form factor $g_{1\Sigma_{c}ND^{*}}$ on Borel
parameter $M_1^2$.\label{your label}}
\end{minipage}
\hfill
\begin{minipage}[t]{0.45\linewidth}
\centering
\includegraphics[height=5cm,width=7cm]{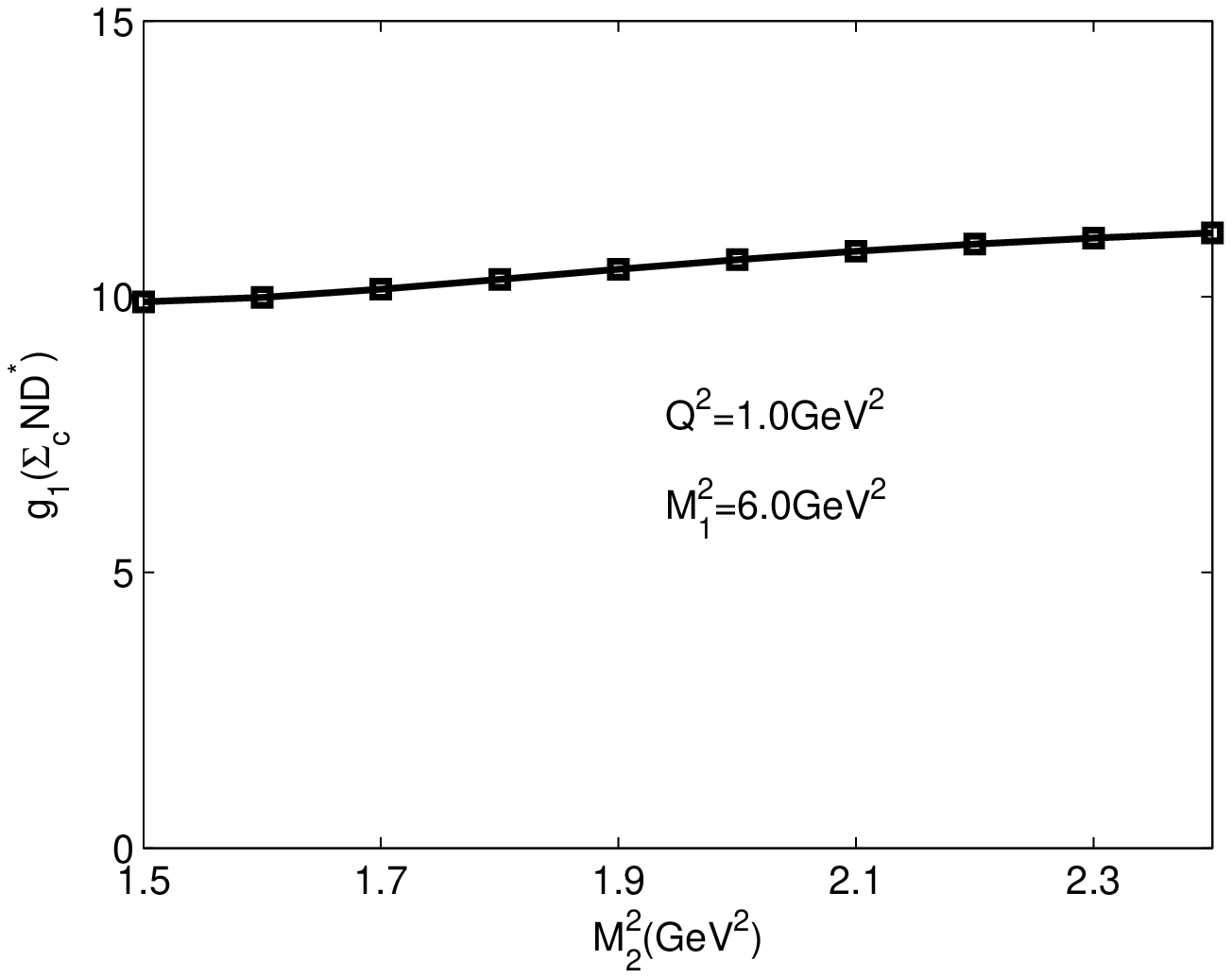}
\caption{The strong form factor $g_{1\Sigma_{c}ND^{*}}$ on Borel
parameter $M_2^2$.\label{your label}}
\end{minipage}
\end{figure}
\begin{figure}[h]
\begin{minipage}[t]{0.45\linewidth}
\centering
\includegraphics[height=5cm,width=7cm]{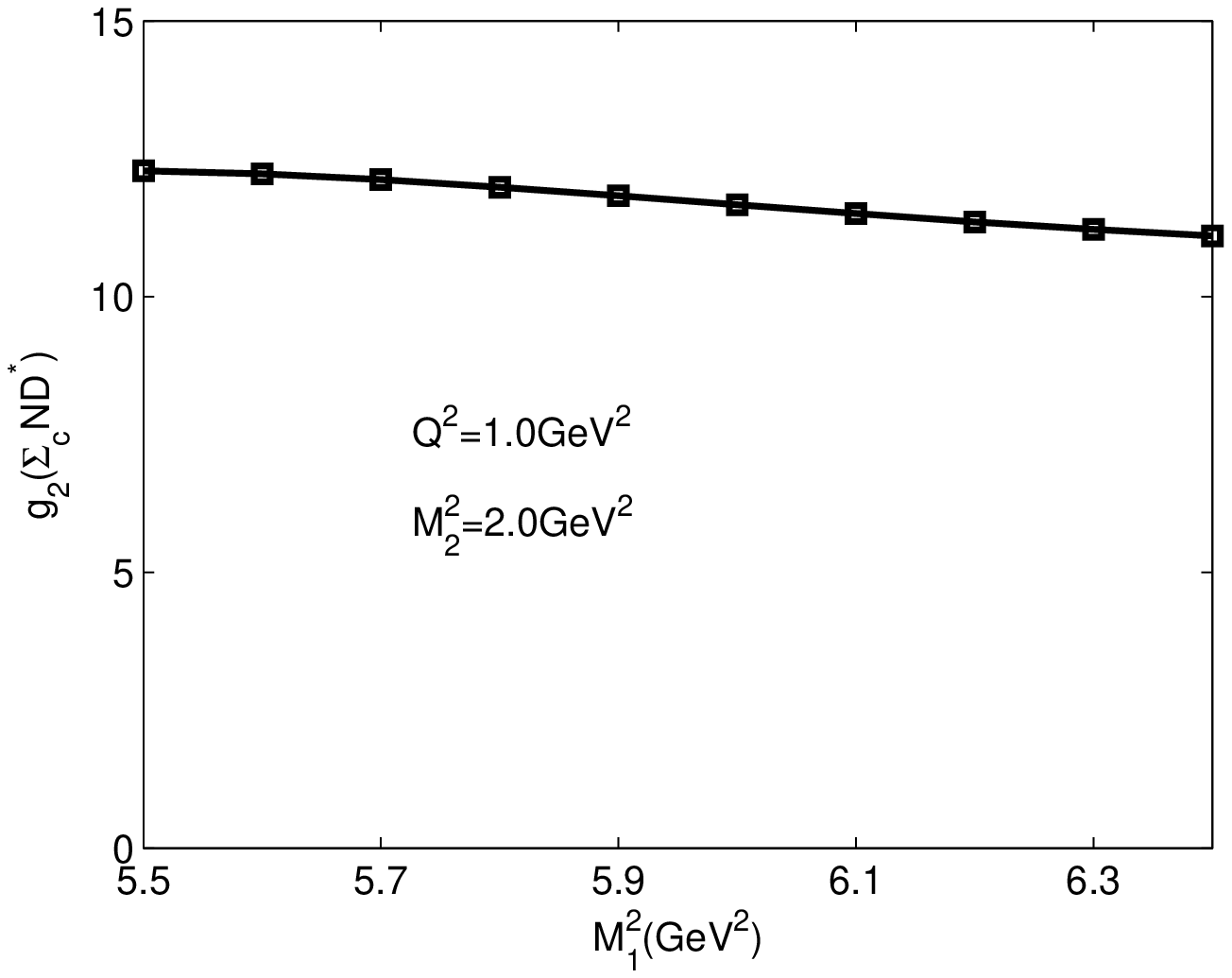}
\caption{The strong form factor $g_{2\Sigma_{c}ND^{*}}$ on Borel
parameter $M_1^2$.\label{your label}}
\end{minipage}
\hfill
\begin{minipage}[t]{0.45\linewidth}
\centering
\includegraphics[height=5cm,width=7cm]{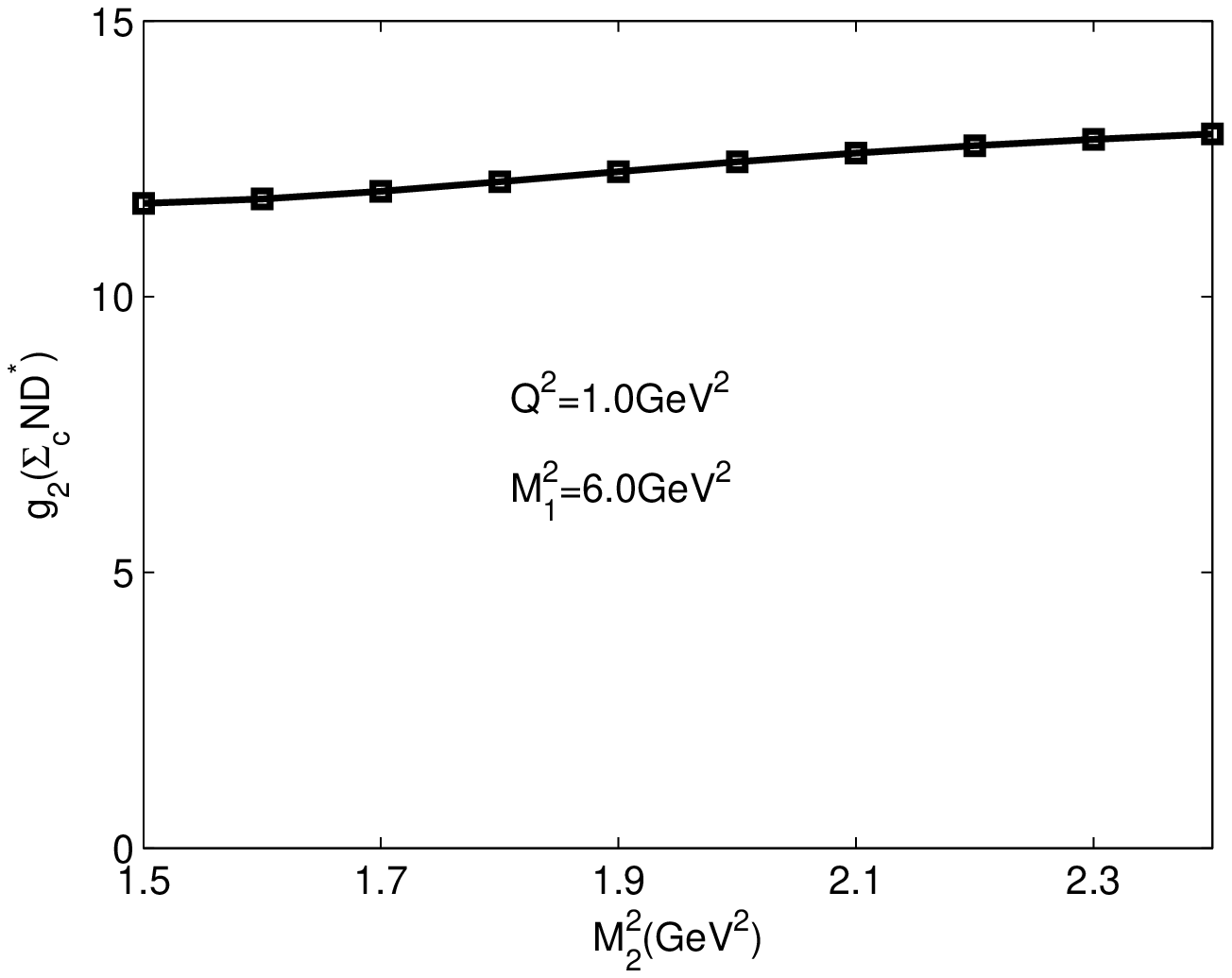}
\caption{The strong form factor $g_{2\Sigma_{c}ND^{*}}$ on Borel
parameter $M_2^2$.\label{your label}}
\end{minipage}
\end{figure}
\begin{figure}[h]
\begin{minipage}[t]{0.45\linewidth}
\centering
\includegraphics[height=5cm,width=7cm]{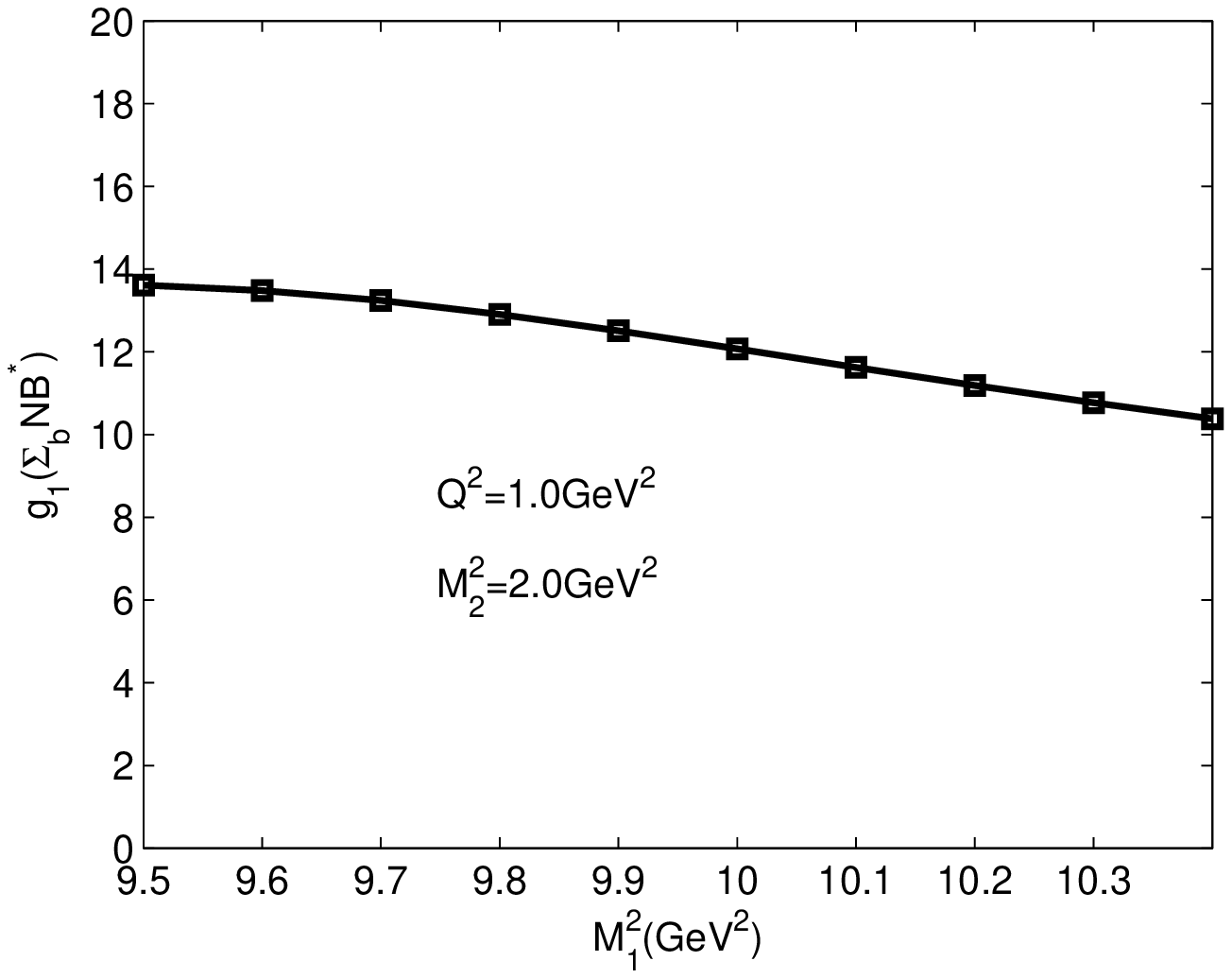}
\caption{The strong form factor $g_{1\Sigma_{b}NB^{*}}$ on Borel
parameter $M_1^2$.\label{your label}}
\end{minipage}
\hfill
\begin{minipage}[t]{0.45\linewidth}
\centering
\includegraphics[height=5cm,width=7cm]{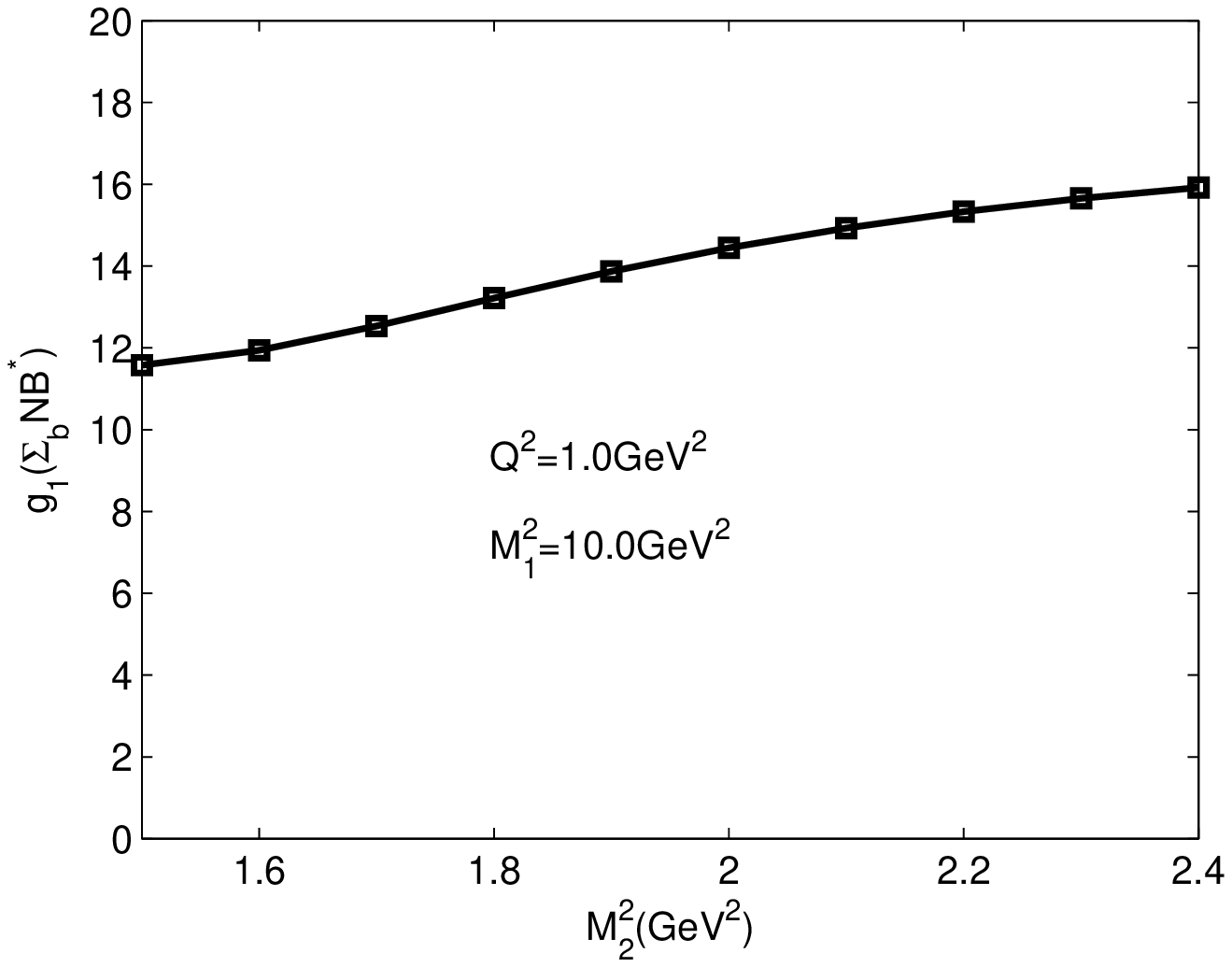}
\caption{The strong form factor $g_{1\Sigma_{b}NB^{*}}$ on Borel
parameter $M_2^2$.\label{your label}}
\end{minipage}
\end{figure}
\begin{figure}[h]
\begin{minipage}[t]{0.45\linewidth}
\centering
\includegraphics[height=5cm,width=7cm]{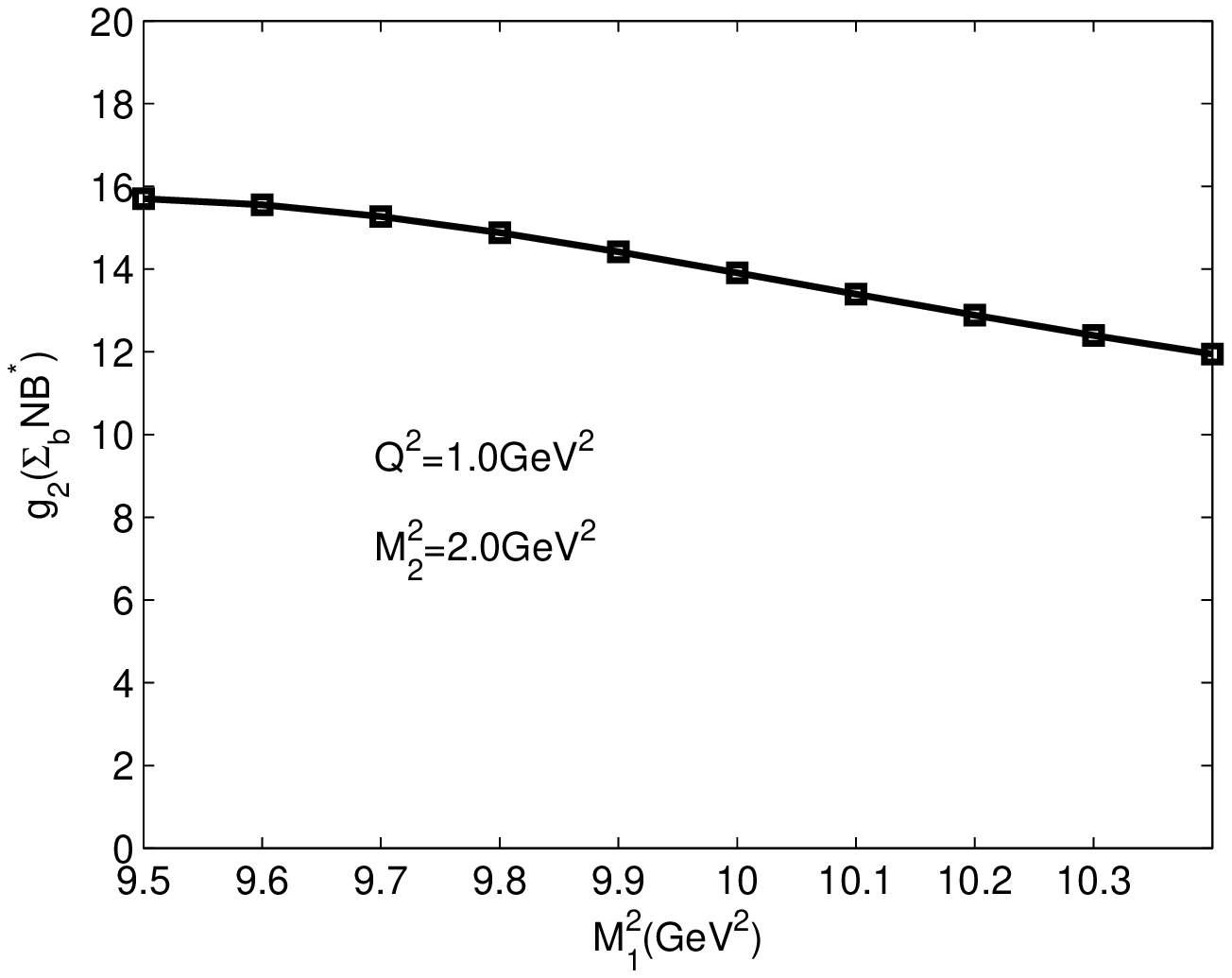}
\caption{The strong form factor $g_{2\Sigma_{b}NB^{*}}$ on Borel
parameter $M_1^2$.\label{your label}}
\end{minipage}
\hfill
\begin{minipage}[t]{0.45\linewidth}
\centering
\includegraphics[height=5cm,width=7cm]{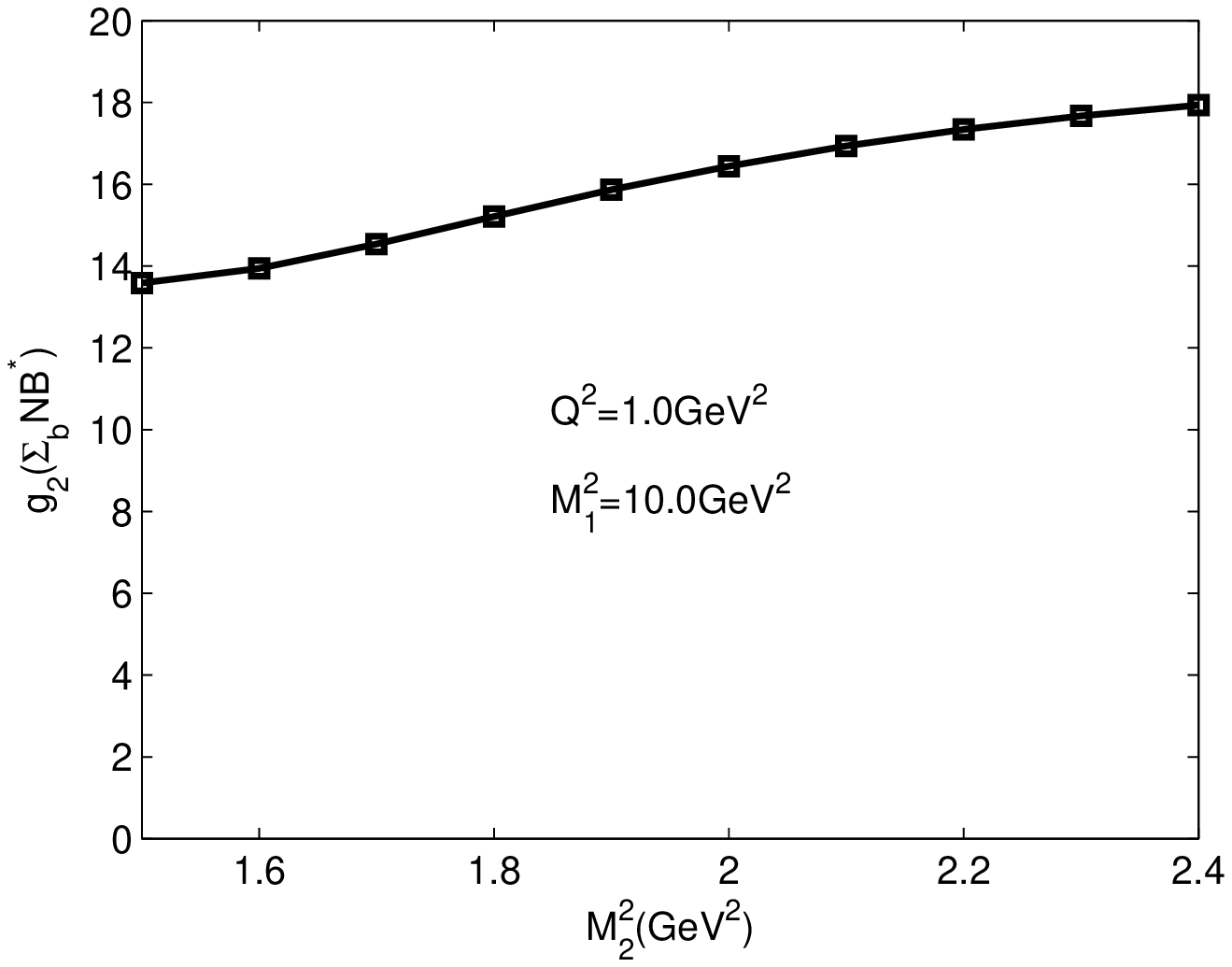}
\caption{The strong form factor $g_{2\Sigma_{b}NB^{*}}$ on Borel
parameter $M_2^2$.\label{your label}}
\end{minipage}
\end{figure}

Physical properties extracted from sum rules must be independent of
Borel parameters $M_1^2$ and $M_2^2$. The assumption is that there
exist a region for these parameters, called Borel window in which
two sides have a overlap and information on the lowest state can be
extracted. Minimum and maximum values for the Borel window can be
determined according to two criterion of QCD sum rules, pole
dominance and OPE convergence. That is to say, pole contribution
should be as large as possible comparing with contributions of
higher and continuum states. Meanwhile, we should also ensure OPE
convergence and the stability of our results. After a comprehensive
consideration, the continuum threshold parameters are chosen to be
$u_{0}=2.07GeV^2$ and $s_{0}=8.72GeV^2[39.90GeV^2]$ for vertex
$\Sigma_{c}ND^{*}[\Sigma_{b}NB]^{*}$.  Besides, the Borel windows
that we choose are listed in Figs.1-8. From these figures we can see
the weak dependence of the results on Borel parameter.

\begin{figure}[h]
\begin{minipage}[t]{0.45\linewidth}
\centering
\includegraphics[height=5cm,width=7cm]{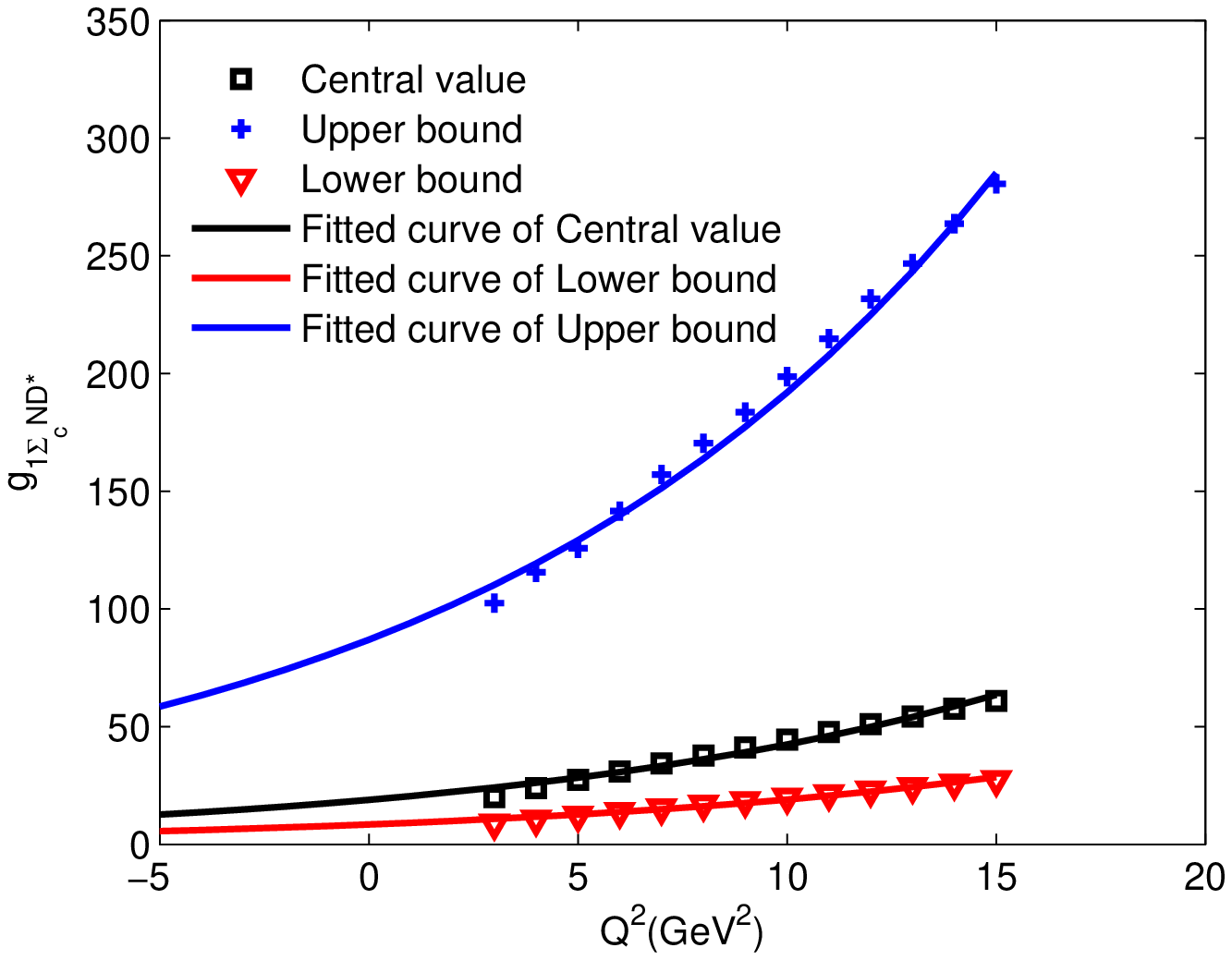}
\caption{The strong form factor $g_{1}$ for the vertex
$\Sigma_{c}ND^{*}$, and its fitted results as a function of
$Q^2$.\label{your label}}
\end{minipage}
\hfill
\begin{minipage}[t]{0.45\linewidth}
\centering
\includegraphics[height=5cm,width=7cm]{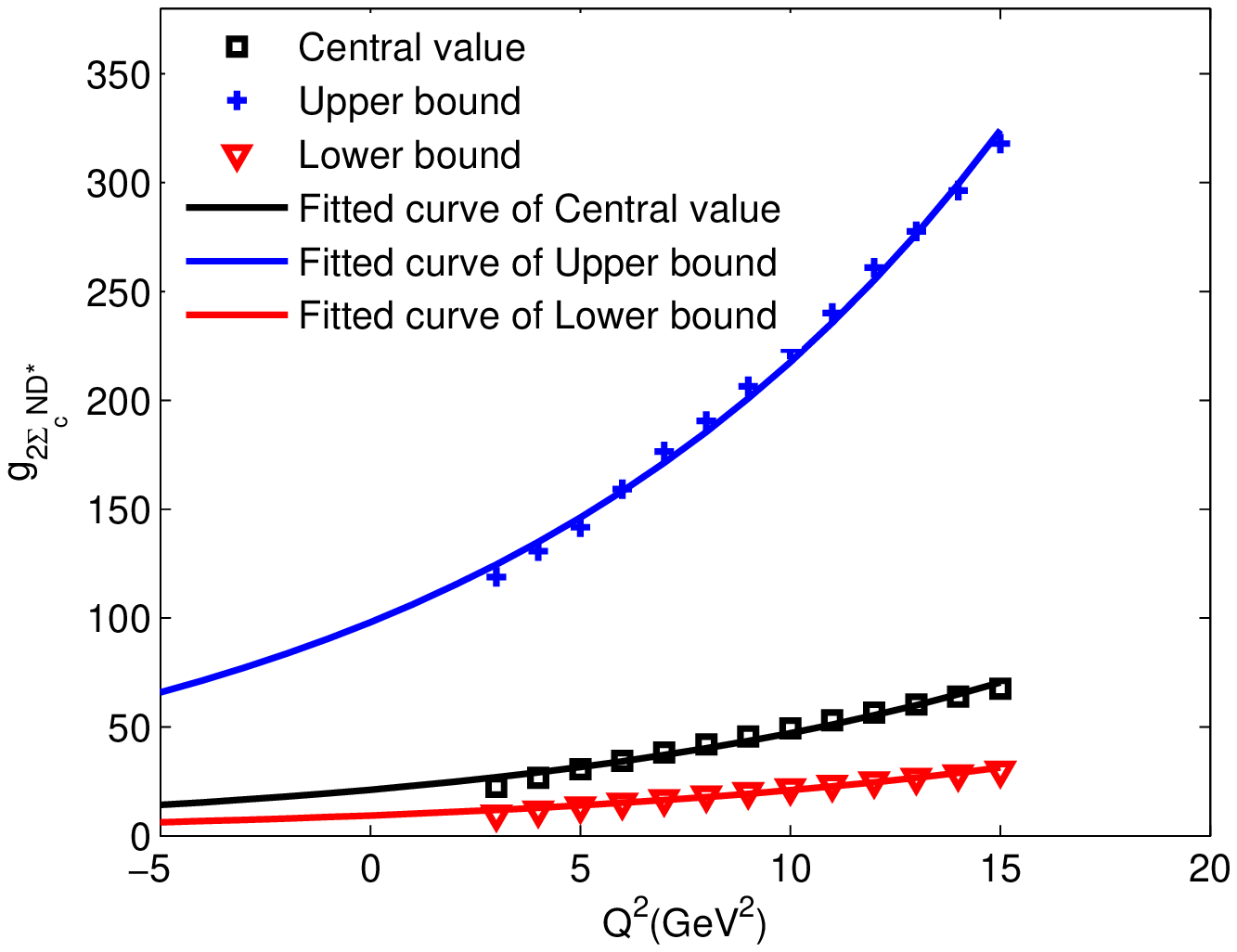}
\caption{The strong form factor $g_{2}$ for the vertex
$\Sigma_{c}ND^{*}$, and its fitted results as a function of
$Q^2$.\label{your label}}
\end{minipage}
\end{figure}

\begin{figure}[h]
\begin{minipage}[t]{0.45\linewidth}
\centering
\includegraphics[height=5cm,width=7cm]{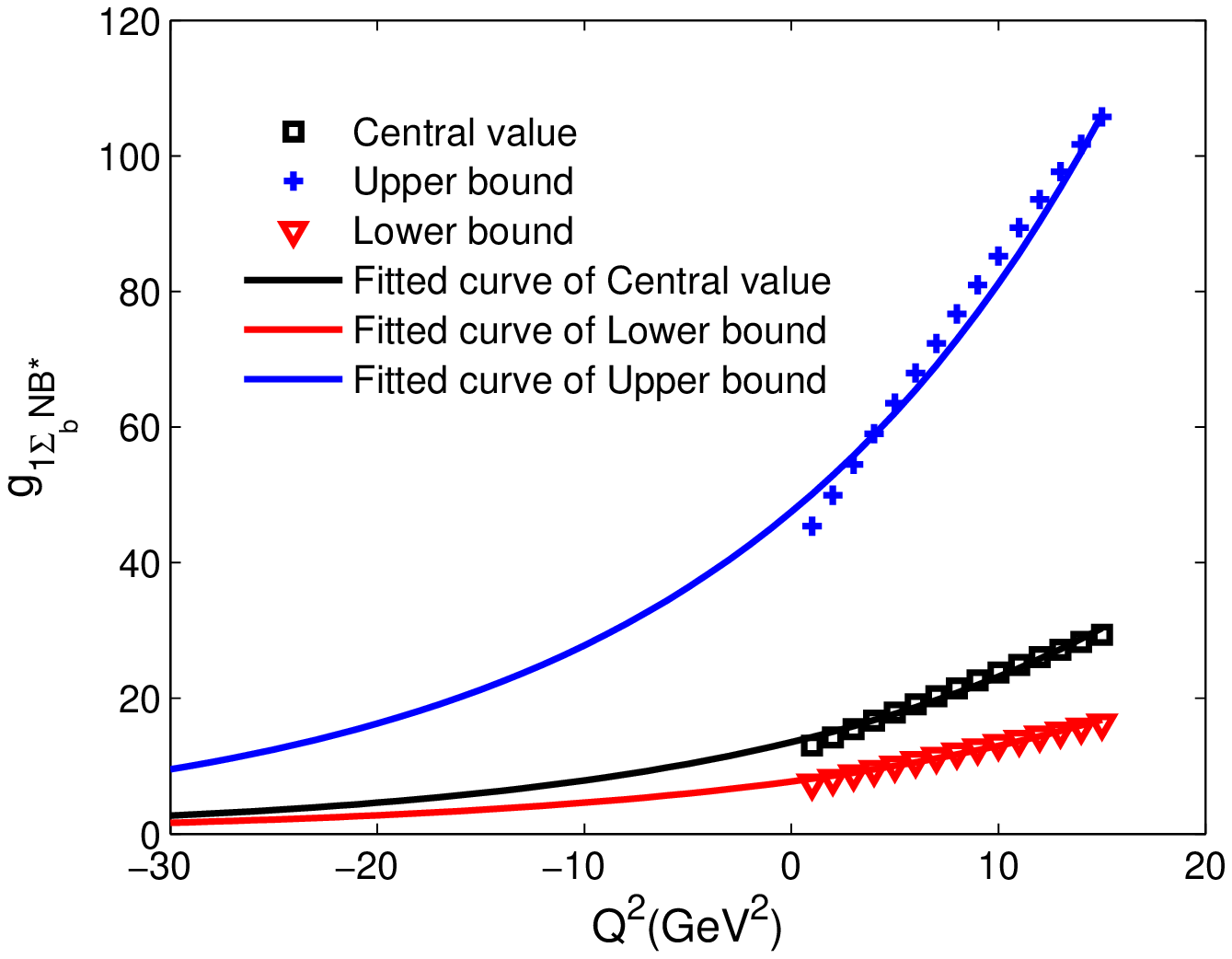}
\caption{The strong form factor $g_{1}$ for the vertex
$\Sigma_{b}NB^{*}$, and its fitted results as a function of
$Q^2$.\label{your label}}
\end{minipage}
\hfill
\begin{minipage}[t]{0.45\linewidth}
\centering
\includegraphics[height=5cm,width=7cm]{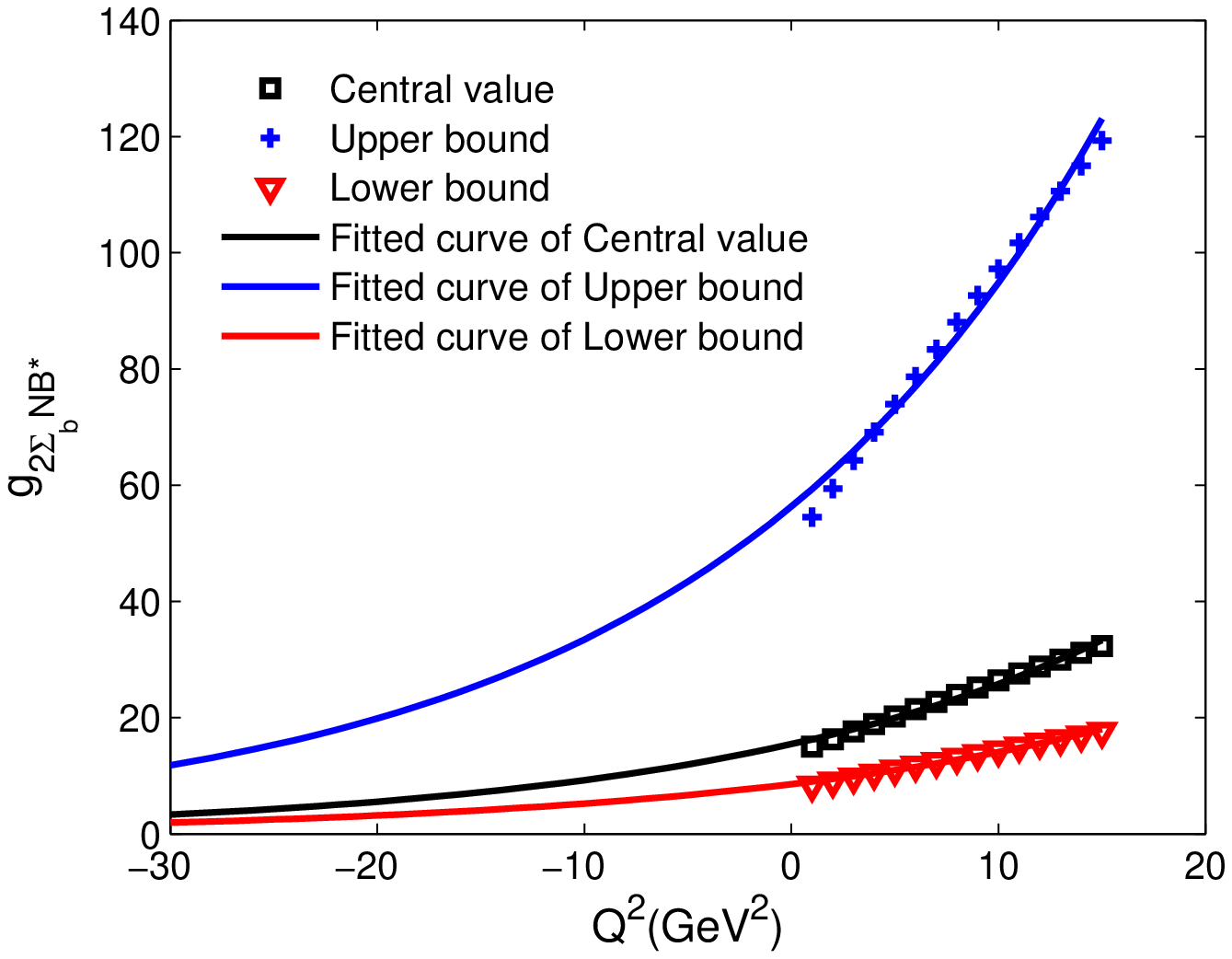}
\caption{The strong form factor $g_{2}$ for the vertex
$\Sigma_{b}^{*}NB$, and its fitted results as a function of
$Q^2$.\label{your label}}
\end{minipage}
\end{figure}

\begin{table*}[t]
\begin{ruledtabular}\caption{Input parameters used in this analysis.}
\begin{tabular}{c c c}
      & \ A   & \ B \\
\hline
$g_{1\Sigma_{c}ND^{*}}$(Q$^2$)    &  \   $18.95\pm3.82$ &  \ $0.08069\pm0.01737$ \\
$g_{2\Sigma_{c}ND^{*}}$(Q$^2$)     &  \    $21.18\pm4.16$   &  \   $0.08014\pm0.01695$   \\
$g_{1\Sigma_{b}NB^{*}}$(Q$^2$)        &  \   $13.52\pm1.2$   &  \ $0.05393\pm0.00843$    \\
$g_{2\Sigma_{b}NB^{*}}$(Q$^2$)       &  \      $15.45\pm1.23$   &  \ $0.05125\pm0.00749$ \\
\end{tabular}
\end{ruledtabular}
\end{table*}

These results are obtained in deep space-like region $q^2
\rightarrow-\infty$, where the intermediate mesons $D^{*}$ and
$B^{*}$ are off-shell. In order to obtain strong coupling constants,
we must extrapolate these results into deep time-like region. This
extrapolation to deep time-like region is mode-dependent, thus there
are no specific expressions for the dependence of the strong form
factors on $Q^2$. Our analysis indicates that this dependence can be
appropriately fitted into the following exponential function,
\begin{equation}
g_{\Sigma_{c}ND^{*}(\Sigma_{b}NB^{*})}(Q^{2})=Aexp[{BQ^2}]
\end{equation}
The fitted results for parameters $A$ and $B$ in this equation are
listed in Table II. In Figs.9-12, we also show the dependence of the
strong form factors on $Q^2$ for the QCD sum rules and the
corresponding fitting results, in which it is marked as Central
value and Fitted curve of Central value. The values of the strong
coupling constants can be obtained from the fitting function at
$Q^{2}=-m_{D^{*}[B^{*}]}^2$, which are
\begin{eqnarray}
\notag\
g_{_{1}\Sigma_{c}ND^{*}}(Q^2=-m_{D^{*}}^{2})=13.69\pm2.92 \\
\notag\
g_{_{2}\Sigma_{c}ND^{*}}(Q^2=-m_{D^{*}}^{2})=15.34\pm3.19 \\
\notag\ g_{_{1}\Sigma_{b}NB^{*}}(Q^2=-m_{B^{*}}^{2})=2.93\pm0.75
\\
\notag\ g_{_{2}\Sigma_{b}NB^{*}}(Q^2=-m_{B^{*}}^{2})=3.61\pm0.82
\end{eqnarray}

The errors appearing in these above results come from the
uncertainties of the fitting parameters $\delta A$ and $\delta B$
which are also listed in Table II. Besides, uncertainties of results
coming from input parameters can theoretically be estimated with
uncertainty transfer formula $\delta=\sqrt{\Sigma_{i}(\frac{\partial
f}{\partial x_{i}})^{2}(x_{i}-\overline{x}_{i})^{2}}$, where $f$
denotes the strong form factors in Eqs.(15) and (16), and $x_{i}$
denotes input parameters
$m_{\Sigma_{b}^{*}}$,$m_{\Sigma_{c}^{*}}$,$m_{b}$,$m_{c}$,$\lambda_{\Sigma_{b}^{*}}$,
$\lambda_{\Sigma_{c}^{*}}$,$\langle \overline{q}q\rangle$,$\cdots$.
For simplicity, the values of the upper and lower limits of the
strong form factors are approximated by taking
$f^{upper(lower)}=f(\overline{x}_{i}\pm\Delta x_{i})$, which are
marked as Upper bound and Lower bound in Figs.9-12. After these
approximations, these results are also fitted into the same kind of
analytical function with Eq.(17) and are also extrapolated into the
physical regions in order to get the uncertainties of the strong
coupling constants. Finally, we obtain the strong coupling
constants,
\begin{eqnarray}
\notag\
g_{_{1}\Sigma_{c}ND^{*}}(Q^2=-m_{D^{*}}^{2})=13.69^{+62.92}_{-6.10}\pm2.92 \\
\notag\
g_{_{2}\Sigma_{c}ND^{*}}(Q^2=-m_{D^{*}}^{2})=15.34^{+71.17}_{-6.77}\pm3.19 \\
\notag\
g_{_{1}\Sigma_{b}NB^{*}}(Q^2=-m_{B^{*}}^{2})=2.93^{+10.08}_{-1.77}\pm0.75
\\
\notag\
g_{_{2}\Sigma_{b}NB^{*}}(Q^2=-m_{B^{*}}^{2})=3.61^{+12.85}_{-2.08}\pm0.82
\end{eqnarray}
where the first part of the uncertainties in the results comes from
the input parameters,
$m_{\Sigma_{b}^{*}}$,$m_{\Sigma_{c}^{*}}$,$m_{b}$,$m_{c}$,$\lambda_{\Sigma_{b}^{*}}$,
$\lambda_{\Sigma_{c}^{*}}$,$\langle \overline{q}q\rangle$,$\cdots$
and the second part originates from the fitting parameters.

\begin{large}
\textbf{4 Conclusion}
\end{large}

In this paper, we perform a systematic analysis on strong vertices
$\Sigma_cND^{*}$ and $\Sigma_bNB^{*}$ with QCD sum rules. We firstly
calculate strong form factors in space-like regions($q^{2}<0$).
Then, the form factors are fitted into analytical functions which
are used to extrapolate into time-like regions($q^{2}>0$) to obtain
strong coupling constants. These results will be valuable for
studying the strong decay behavior of the charmed and bottom baryons
in the future.

%\end{CJK*}

\begin{large}
\textbf{Acknowledgment}
\end{large}

This work has been supported by the Fundamental Research Funds for
the Central Universities, Grant Number $2016MS133$, Natural Science
Foundation of HeBei Province, Grant Number $A2018502124$.

\end{document}